\documentclass{llncs}
\usepackage{amsmath,amssymb}
\usepackage[all]{xy}
\usepackage{euscript}
\usepackage{anysize}
\marginsize{3cm}{3cm}{2cm}{2cm}

\title{Modules over Monads and Linearity}
\author{Andr\'e Hirschowitz\inst{1} \and Marco Maggesi\inst{2}}

\institute{LJAD, Universit\'e de Nice Sophia--Antipolis, CNRS\\
\texttt{http://math.unice.fr/\homedir ah}
\and Universit\`a degli Studi di Firenze\\
\texttt{http://www.math.unifi.it/\homedir maggesi}}

\DeclareMathOperator{\Mon}{Mon}
\DeclareMathOperator{\Mod}{Mod}

\newcommand{\coloneq}{\mathrel{\mathop:}=}

\newcommand{\NN}{\mathbb{N}}
\newcommand{\Cat}[1]{\mathsf{#1}}
\newcommand{\CC}{\Cat{C}}
\newcommand{\DD}{\Cat{D}}

\newcommand{\LC}{{\mathsf{LC}}}
\newcommand{\Set}{{\mathsf{Set}}}
\newcommand{\var}{{\mathsf{var}}}
\newcommand{\app}{{\mathsf{app}}}
\newcommand{\abs}{{\mathsf{abs}}}
\newcommand{\fold}{{\mathsf{fold}}}
\newcommand{\bind}{{\mathsf{bind}}}
\newcommand{\mbind}{{\mathsf{mbind}}}
\newcommand{\Maybe}{{\mathsf{Maybe}}}
\newcommand{\nil}{{\mathsf{nil}}}
\newcommand{\cons}{{\mathsf{cons}}}

\begin{document}
\maketitle
\begin{abstract}
  Inspired by the classical theory of modules over a monoid, we give a
  first account of the natural notion of module over a monad.  The
  associated notion of morphism of left modules ("linear" natural
  transformations) captures an important property of compatibility
  with substitution, in the heterogeneous case where "terms" and
  variables therein could be of different types as well as in the
  homogeneous case.  In this paper, we present basic constructions of
  modules and we show examples concerning in particular abstract
  syntax and lambda-calculus.
\end{abstract}

\section {Introduction}
\label{sec:intro}

Substitution is a major operation.  Its relevance to computer sciences
has been stressed constantly (see e.g.\
\cite{Fiore06}). Mathematicians of the last century have coined two
strongly related notions which capture the formal properties of this
operation. The first one is the notion of monad, while the second one
is the notion of operad. We focus on the notion of monad. A monad in
the category $\CC$ is a monoid in the category of endofunctors of
$\CC$ (see \ref{sec:monads} below) and as such, has right and left
modules. Apriori these are endofunctors (in the same category) equipped
with an action of the monad. In fact, we introduce a slightly more general
notion of modules over a monad, based on the elementary observation
that we can readily extend the notion of a right action of a monad in
$\CC$ to the case of a functor from any category $\Cat{B}$ to $\CC$,
and symmetrically the notion of a left action of a monad in $\CC$ to
the case of a functor from $\CC$ to any category $\DD$.  We are mostly
interested in left modules. As usual, the interest of the notion of
left module is that it generates a companion notion of morphism. We
call morphisms those natural transformations among (left) modules
which are compatible with the structure, namely which commute to
substitution (we also call these morphisms {\em linear} natural
transformations).

Despite the natural ideas involved, the only mention of modules over
monads we have been able to find is on a blog by Urs
Schreiber.\footnote{http://golem.ph.utexas.edu/string/archives/000715.html}
On the other hand, modules over operads have been introduced by
M.~Markl (\cite{arXiv:hep-th/9411208v1,arXiv:hep-th/9608067v1}) and
are commonly used by topologists (see
e.g.~\cite{MR2066499,arXiv:math/0607427v1,MR2140994}).  In \cite{FT},
such modules over operads have been considered, under the name of
actions, in the context of semantics.

We think that the notions of module over a monad and linear
transformations deserve more attention and propose here a first
reference for basic properties of categories of left modules, together
with basic examples of morphisms of left modules, hopefully showing
the adequacy of the language of left modules for questions concerning
in particular (possibly higher-order) syntax and lambda-calculus.

In section 2, we briefly review the theory of monads and their
algebras. In section 3, we develop the basic theory of modules. In
section 4, we sketch a treatment of syntax (with variable-binding)
based on modules.  In the remaining sections, we show various linear
transformations concerning lists and the lambda-calculus (typed or
untyped).  The appendix discusses the formal proof in the Coq proof
assistant of one of our examples.

\section{Monads and Algebras}
\label{sec:monads}

We briefly recall some standard material about monads and algebras.
Experienced readers may want to skip this section or just use it as
reference for our notations.  Mac Lane's book \cite{maclane} can be
used as reference on this material.

Let $\CC$ be a category.  A monad over $\CC$ is a monoid in the
category $\CC\to\CC$ of endofunctors of $\CC$.  In more concrete
terms:
\begin{definition}[Monad]
  A monad $R=\langle R,\mu,\eta \rangle$ is given by a functor
  $R\colon\CC \to \CC$, and two natural transformations $\mu\colon R^2
  \to R$ such that the following diagrams
  commute:
  \begin{equation*}
    \xymatrix{
      R^3 \ar[d]_{\mu R} \ar[r]^{R\mu} & R^2 \ar[d]^{\mu} \\
      R^2 \ar[r]^{\mu} & R}\qquad
    \xymatrix{
      I\cdot R \ar[rd]_{1_R} \ar[r]^{\eta R} & R^2 \ar[d]^{\mu} &
      R\cdot I \ar[ld]^{1_R} \ar[l]_{R \eta}  \\
      {} & R {}}
  \end{equation*}
  The $\mu$ and $\eta$ natural transformations are often referred as
  \emph{product} (or \emph{composition}) and \emph{unit} of the monad
  $M$.  In the programming language Haskell, they are noted
  $\mathtt{join}$ and $\mathtt{return}$ respectively.
\end{definition}

Given a monad $R$ and an arrow $f\colon X\to R Y$, we define the
function $\bind\,f\colon RX \to RY$ given by $\bind\,f \coloneq \mu
\cdot Rf$.  The functoriality and the composition of the monad can be
defined alternatively in terms of the unit and the $\bind$ operator.
More precisely, we have the equations
\begin{equation*}
  \mu_X = \bind\,1_X,\qquad R f = \bind (\eta\cdot f).
\end{equation*}
Moreover, we have the following associativity and unity equations for
$\bind$
\begin{equation}
  \label{eq:bind-ax}
  \bind\,g \cdot \bind\,f = \bind (\bind\,g\cdot f), \quad
  \bind\,\eta_X = 1_{RX},\quad
  \bind\,f \cdot \eta = f
\end{equation}
for any pair of arrows $f\colon X \to R Y$ and $g\colon Y\to R Z$.

In fact, to give a monad is equivalent to give two operators unit and
bind as above which satisfy equations \ref{eq:bind-ax}.

\begin{example}[Lists]
  To construct the monad of lists $L$ (over $\Set$), first take the
  functor $L\colon\Set \to \Set$
  \begin{equation*}
    L \colon X \mapsto \Sigma_{n\in\mathbb{N}} X^n =
    * + X + X\times X + X\times X\times X + \dots
  \end{equation*}
  So $L(X)$ is the set of all finite lists with elements in $X$.  Then
  consider as composition the natural transformation $\mu\colon L\cdot
  L \to L$ given by the \emph{join} (or \emph{flattening}) of lists of
  lists:
  \begin{equation*}
    \mu [[a_1, \dots], [b_1,\dots], \dots, [z_1,\dots]]
    = [a_1,\dots, b_1,\dots, \dots, z_1,\dots].
  \end{equation*}
  The unit $\eta\colon I \to L$ is constituted by the singleton map
  $\eta_X \colon x\in X\mapsto [x]\in L(X)$.
\end{example}

\begin{example}[Lambda Calculus]
  This example will be worked out with the necessary details in
  section \ref{sec:untyp-lambda-calc}, but let us give early some
  basic ideas (see also \cite{Alt-Reus}).  We denote by
  $\mathop{FV}(M)$ the set of free variables of a $\lambda$-term $M$.
  For a fixed set $X$, consider the collection of $\lambda$-terms (modulo
  $\alpha$-conversion) with free variables in $X$:
  \begin{equation*}
    \LC(X) \coloneq \{ M | \mathop{FV}(M) \subset X \}.
  \end{equation*}
  Given a set $X$ we take as unit morphism $\eta_X \colon X \to
  \LC(X)$ the application assigning to an element $x \in X$ the
  corresponding variable in $\LC(X)$.  Every map $f \colon X \to Y$
  induces a morphism $\LC(f)\colon \LC(X) \to \LC(Y)$ (``renaming'')
  which makes $\LC$ a functor.  The instantiation (or substitution) of
  free variables gives us a natural transformation
  \begin{equation*}
    \mu_X \colon \LC(\LC(X)) \to \LC(X).
  \end{equation*}
  With this structure $\LC$ is a monad.

  Moreover, by taking the quotient $\Lambda(X)$ of $\LC(X)$ modulo
  $\beta\eta$-conversion we still obtain a monad (i.e., the
  composition and the unit of the monad are compatible with
  $\beta\eta$-conversions).
\end{example}

\begin{definition}[$\Maybe$ monad]
  In a category $\CC$ with finite sums and a final object (like
  $\Set$), the functor $X\mapsto X+*$ which takes an object and ``adds
  one point'' has a natural structure of monad on $\CC$.  Borrowing
  from the terminology of the library of the programming language
  Haskell, we call it the $\mathtt{Maybe}$ monad.
\end{definition}

\begin{definition}[Derivative]
  We define the \emph{derivative} $F'$ of a functor $F\colon \CC \to
  \CC$ to be the functor $F'=F\cdot \Maybe$.  We can iterate the
  construction and denote by $F^{(n)}$ the $n$-th
  derivative.\footnote{This corresponds to the $\mathtt{MaybeT}$ monad
    transformer in Haskell.}
\end{definition}

\begin{definition}[Morphisms of monads]
  A morphism of monads is a natural transformation between two monads
  $\phi\colon P\to R$ which respects composition and unit, i.e., such
  that the following diagrams commute:
  \begin{equation*}
    \xymatrix{
      P^2 \ar[d]^{\mu_R} \ar[r]^{\phi\phi} & R^2 \ar[d]^{\mu_R} \\
      P \ar[r]_{\phi} & R}\qquad
    \xymatrix{
      P\ar[rr]^\phi & & R \\
      {} & I\ar[ul]^{\eta_P}\ar[ur]_{\eta_r}}
  \end{equation*}
  It can be easily seen that morphisms of monads form a category.
\end{definition}

For our purpose it is relevant to observe that there are a number of
natural transformations which arise in the above examples which fail
to be morphisms of monads.  We take the following as paradigmatic
example.
\begin{example}[Abstraction is not a morphism of monads]
  Abstraction on $\lambda$-terms gives a natural transformation
  $\abs\colon \LC'\to \LC$ which takes a $\lambda$-term $M\in\LC(X+*)$
  and binds the ``variable'' $*$.  This fails to be a morphism of
  monads because it does not respect substitution in the sense of
  monads: a careful inspection reveals that the transformation
  \begin{equation*}
    \LC(\LC(X+*)+*) \stackrel{\mu}\longrightarrow
    \LC(X+*) \stackrel{\abs}\longrightarrow \LC(X)
  \end{equation*}
  binds all stars under a single abstraction while
  \begin{equation*}
    \LC(\LC(X+*)+*)
    \stackrel{\abs\,\abs}\longrightarrow \LC(\LC(X))
    \stackrel{\mu}\longrightarrow \LC(X)
  \end{equation*}
  not.  In fact, we will see later that $\LC'$ is a left module over
  $\LC$ and $\abs$ is a $\LC$-\emph{linear} morphism.
\end{example}

Now let $R$ be a monad over $\CC$.

\begin{definition}[Algebra]
  An \emph{algebra} over $R$ is given by an object $A$ and a morphism
  $\rho\colon R(A) \to A$ in $\CC$ such that the following diagrams
  commute:
  \begin{equation*}
    \xymatrix{
      R^2(A) \ar[d]_{\mu_A} \ar[r]^{R\rho} & R(A) \ar[d]^{\rho} \\
      R(A) \ar[r]^{\rho} & A}\qquad
    \xymatrix{
      A \ar[rd]_{1_A} \ar[r]^{\eta_A} & R(A) \ar[d]^{\rho} \\
      {} & A}
  \end{equation*}
\end{definition}

\begin{definition}
  Let $A$, $B$ be two algebras over a monad $R$.  An arrow $f\colon
  A\to B$ in $\CC$ is said to be a morphism of algebras if it is
  compatible with the associated actions, i.e., the two induced
  morphisms from $R(A)$ to $B$ are equal:
  \begin{equation*}
    \rho_B\cdot R(f) = f\cdot \rho_A
  \end{equation*}
\end{definition}

As we will see later, algebras can be regarded as special kind of
right modules.

\begin{example}[Monoids]
  In the category of sets, algebras over the monad $L$ of lists are
  sets equipped with a structure of monoid; given a monoid $A$, the
  corresponding action $L(A) \to A$ is the product (sending a list to
  the corresponding product).
\end{example}

\section {Modules over monads}
\label{sec:modules}

Being a monoid in a suitable monoidal category, a monad has associated
left and right modules which, a-priori, are objects in the same
category, acted upon by the monoid.

Although we are mostly interested in left modules, let us remark that
from this classical point of view, algebras over a monad are not
(right-)modules.  We give a
slightly more general definition of modules which is still completely
natural.  According to this extended definition, algebras turn out to
be right-modules.

\subsection{Left modules}

We start first by concentrating ourselves on left modules over a
given monad $R$ over a category $\CC$.

\begin{definition}[Left modules]
  A left $R$-module in $\DD$ is given by a functor $M\colon \CC \to \DD$
  equipped with a natural transformation $\rho\colon M \cdot R \to M$,
  called \emph{action}, which is compatible with the monad
  composition, more precisely, we require that the following diagrams
  commute
  \begin{equation*}
    \xymatrix{
      M\cdot R^2 \ar[d]_{\rho R} \ar[r]^{M\mu} & M\cdot R \ar[d]^{\rho} \\
      M\cdot R \ar[r]^{\rho} & M}\qquad
    \xymatrix{
      M\cdot R \ar[d]^{\rho} &
      M\cdot I \ar[ld]^{1_M} \ar[l]_{M \eta}  \\
      M}
  \end{equation*}
  We will refer to the category $\DD$ as the \emph{range} of $M$.
\end{definition}

\begin{remark}
  The companion definition of modules over an operad
  (c.f.~e.g.~\cite{arXiv:hep-th/9608067v1,MR2066499}) follows easily
  from the observation \cite{smirnov86} that operads are monoids in a
  suitable monoidal category. This monoidal structure is central in
  \cite{FPT}.
\end{remark}

Given a left $R$-module $M$, we can introduce the $\mbind$ operator
which, to each arrow $f\colon X\to R Y$, associates an arrow $\mbind\,f
\coloneq M X \to M Y$ defined by $\mbind \,f \coloneq \rho\cdot Mf$.
The axioms of left module are equivalent to the following equations
over $\mbind$:
\begin{equation*}
  \label{eq:mbind-ax}
  \mbind\,g \cdot \mbind\,f = \mbind (\bind\,g\cdot f), \qquad
  \mbind\,\eta_X = 1_X
\end{equation*}

\begin{example}
  We can see our monad $R$ as a left module over itself (with range
  $\CC$), which we call the \emph{tautological} module.
\end{example}

\begin{example}
  Any constant functor $\underline W \colon \CC \to \DD$, $W\in \DD$
  is a trivial left $R$-module.
\end{example}

\begin{example}
  For any functor $F\colon \DD \to \Cat{E}$ and any left $R$-module
  $M\colon \CC \to \DD$, the composition $F\cdot M$ is a left
  $R$-module (in the evident way).
\end{example}

\begin{definition}[Derived module]
  As for functors and monads, derivation is well-behaved also on left
  modules: for any left $R$-module $M$, the derivative $M'=M\cdot
  \Maybe$ has a natural structure of left $R$-module where the action
  $M'\cdot P \to M'$ is the composition
  \begin{equation*}
    M\cdot\Maybe\cdot R \stackrel{M\gamma}\longrightarrow
    M\cdot R\cdot \Maybe \stackrel{\rho\Maybe}\longrightarrow
    M\cdot\Maybe
  \end{equation*}
  and $\gamma$ is the natural arrow $\Maybe\cdot R \to R\cdot\Maybe$.
\end{definition}

\begin{definition}[Morphisms of left modules]
  We say that a natural transformation of left $R$-modules
  $\tau\colon M \to N$ is \emph{linear} if it is compatible with
  substitution:
  \begin{equation*}
    \xymatrix{
      M\cdot R \ar[d]_{\rho_M} \ar[r]^{\tau R} & N\cdot R \ar[d]^{\rho_N} \\
      M \ar[r]^{\tau} & N}
  \end{equation*}
  We take linear natural transformations as left module morphisms.
\end{definition}

\begin{remark}
  Here the term {\em linear} refers to linear algebra: linear
  applications between modules over a ring are group morphisms
  compatible with the action of the ring.  It is compatible with the
  usual flavor of the word linear (no duplication, no junk) as the
  following example shows.
\end{remark}

\begin{example}
  We consider the monad $M$ on $\Set$ generated by two binary
  constructions $+$ and $*$ and we build (by recursion) a natural
  transformation $n\colon M \to M$ as follows: for a variable $x$,
  $n(x)$ is $x+x$, while for the other two cases we have
  $n(a+b)=n(a)*n(b)$ and $n(a*b)=n(a)+n(b)$.  It is easily verified
  that $n$ is a non-linear natural transformation (check the diagram
  against $n(\var(x*x))$).
\end{example}

\begin{example}
  We easily check that the natural transformation of a left module
  into its derivative is linear.  Note that there are two natural
  inclusions of the derivative $M'$ into the second derivative $M''$.
  Both are linear.
\end{example}

\begin{example}
  Consider again the monad of lists $L$.  The concatenation of two
  lists is a $L$-linear morphism $L\times L\to L$.
\end{example}

\begin{definition}[Category of left modules]
  We check easily that linear morphisms between left $R$-modules with
  the same range yield a subcategory of the functor category.  We
  denote by $\Mod^\DD(R)$ the category of left $R$-modules with range
  $\DD$.
\end{definition}

\begin{definition}[Product of left modules]
  We check easily that the cartesian product of two left $R$-modules
  as functors (having as range a cartesian category $\DD$) is
  naturally a left $R$-module again and is the cartesian product also
  in the category $\Mod^\DD(R)$.  We also have finite products as
  usual.  The final left module $*$ is the product of the empty
  family.
\end{definition}

\begin{example}
  Given a monad $R$ on $\Set$ and a left $R$-module $M$ with range in
  a fixed cartesian category $D$, we have a natural ``evaluation''
  morphism
  \begin{math}
    \mathsf{eval} \colon M' \times R \to M
  \end{math},
  where $M'$ is the derivative of $M$.
\end{example}

\begin{proposition}
  Derivaton yields a cartesian endofunctor on the category of left
  $R$-modules with range in a fixed cartesian category $D$
\end{proposition}

\subsection{Right modules}

Let $R$ be a monad over a category $\CC$.  The definition of right
module is similar to that of left module.

\begin{definition}[Right modules]
  A right $R$-module (from $\DD$) is given by a functor $M\colon \DD \to \CC$
  equipped with a natural transformation $\rho\colon R \cdot M \to M$
  which makes the following diagrams commutative
  \begin{equation*}
    \xymatrix{
      R^2\cdot M \ar[d]_{R\rho} \ar[r]^{\mu M} & R\cdot M \ar[d]^{\rho} \\
      R\cdot M \ar[r]^{\rho} & M}\qquad
    \xymatrix{
      I\cdot M \ar[rd]_{1_M} \ar[r]^{\eta M} &
      R\cdot M \ar[d]^{\rho} \\
      & M}
  \end{equation*}
  As for left modules, we will call \emph{corange} of $M$ the category
  $\DD$.
\end{definition}

We remark that for any right $R$-module $M$ and any object $X\in\DD$
the image $M(X)$ is an $R$-algebra.  Then a right $R$-module is simply a
functor from the corange category $\DD$ to the category of $R$-algebras.

\begin{example}
  Our monad $R$ is a right module over itself.
\end{example}

\begin{example}
  If $A$ is an $R$-algebra, then for any category $\DD$ the constant
  functor $\underline A \colon X \mapsto A$ has a natural structure of
  right $R$-module.  In particular, we can identify our algebra with
  the corresponding functor $\underline A\colon * \to \CC$, where $*$
  is the category with one object and one arrow.
\end{example}

\begin{example}
  Let $\phi\colon R\to P$ be a morphism of monads.  Then $P$ is a right
  and left $R$-module with actions given respectively by
  \begin{math}
    P\cdot R \stackrel{\phi R}\longrightarrow R\cdot R
    \stackrel{\mu_R}\longrightarrow R
  \end{math}
  and
  \begin{math}
    R\cdot P \stackrel{P\phi}\longrightarrow R\cdot R
    \stackrel{\mu_R}\longrightarrow R
  \end{math}.
\end{example}

\begin{definition}[Morphisms of right modules]
  A morphism of right $R$-modules is a natural transformation $\tau
  \colon M \to N$ which is compatible with substitution, i.e., such
  that the following diagram commutes:
  \begin{equation*}
    \xymatrix{
      R\cdot M \ar[d]_{\rho_M} \ar[r]^{R \tau} & R\cdot N \ar[d]^{\rho_N} \\
      M \ar[r]^{\tau} & N}
  \end{equation*}
\end{definition}

\begin{definition}[Category of right $R$-modules]
  We check easily that module morphisms among right $R$-modules with
  the same corange yield a subcategory of the functor category.
\end{definition}

\subsection{Limits and colimits of left modules}

Limits and colimits in the category of left modules can be
constructed pointwise.  For instance:
\begin{lemma}[Limits and colimits of left modules]
  If $\DD$ is complete (resp.\ cocomplete), then $\Mod^\DD(R)$ is
  complete (resp.\ cocomplete).
\end{lemma}
\begin{proof}
  Suppose first that $\DD$ be a complete category and $G\colon \Cat{I}
  \to \Mod^\DD(R)$ be a diagram of left modules over the index
  category $\Cat{I}$.  For any object $X\in \CC$ we have a diagram
  $G(X) \colon \Cat{I} \to \DD$ and for any arrow $f\colon X \to Y$ in
  $\CC$ we have a morphism of diagrams $G(X) \to G(Y)$.  So define
  \begin{equation*}
    U(X) \coloneq \lim G(X)
  \end{equation*}
  Next, given an arrow $f\colon X\to R(Y)$, we have an induced
  morphism of diagrams $G(X)\to G(Y)$ by the module structure on each
  object of the diagram.  This induces a morphism $\mbind\,f \colon
  U(X)\to U(Y)$.  It is not hard to prove that $\mbind$ satisfies the
  module axioms and that $U$ is the limit of $G$.  The colimit
  construction is carried analogously.
\end{proof}

\subsection{Base change}

\begin{definition}[Base change]
  Given a morphism $f\colon A\to B$ of monads and a left $B$-module
  $M$, we have an $A$-action on $M$ given by
  \begin{equation*}
    M\cdot A\stackrel{Mf}{\longrightarrow} M\cdot B
    \stackrel{\rho_M}\longrightarrow M.
  \end{equation*}
  We denote by $f^*\!M$ the resulting $A$-module and we refer to $f^*$
  as the \emph{base change} operator.
\end{definition}

\begin{lemma}
  The base change of a left module is a left module.
\end{lemma}
\begin{proof}
  Our thesis is the commutativity of the diagram
  \begin{equation*}
    \xymatrix{
      M\cdot B\cdot A \ar[dd]_{\rho A} & M\cdot A\cdot
      A\ar[l]_{MfA}\ar[d]_{Mff}\ar[r]^{M\mu}
      & M\cdot A\ar[d]^{Mf}\\
      {} & M\cdot B\cdot B\ar[d]_{\rho B}\ar[r]^{M\mu}
      & M\cdot B\ar[d]^\rho \\
      M\cdot A\ar[r]_{Mf} & M\cdot B \ar[r]_\rho & M}
  \end{equation*}
  which follows from the commutativity of the three pieces: $M$
  is a left $B$-module, the map from $M(B(\_))\to M(\_)$ is functorial,
  and $f$ is a morphism.
\end{proof}

\begin{definition}[Base change (functoriality)]
  We upgrade the base change operator into a functor
  $f^*\colon \Mod^\DD(B) \to \Mod^\DD(A)$
  by checking that if $g\colon M \to N$ is a morphism of
  left $B$-modules, then so is the natural transformation $f^*g
  \colon f^*M \to f^* N$.
\end{definition}

\begin{proposition}
  \label{prop:pull-back-commute}
  The base change functor commutes with products and with derivation.
\end{proposition}

\begin{proposition}
  Any morphism of monads $f\colon A \to B$ yields a morphism of left
  $A$-modules, still denoted $f$, from $A$ to $f^*B$.
\end{proposition}

\section {Initial Algebra Semantics}
\label{sec:semantics}

To ease the reading of the forthcoming sections, we collect in this
section some classical ideas about Initial Algebra Semantics.

Given a category $\CC$ and an endofunctor $T\colon \CC \to \CC$, a
$T$-algebra\footnote{There is a lexical conflict here with algebra of
  monads introduced in section \ref{sec:monads}, which is deeply
  rooted in the literature anyway.  We hope that this will not lead to
  any confusion.}  is given by an object $A\in\CC$ and an arrow
\begin{equation*}
  \sigma_A\colon T A \to A.
\end{equation*}
A morphism of $T$-algebras is an arrow
$f:A\to B$ which commutes with the \emph{structural morphism} $\sigma$
\begin{equation*}
  \xymatrix{
    T A \ar[d]_{\sigma_A} \ar[r]^{T f} &  T B\ar[d]^{\sigma_B} \\
    A \ar[r]_f & B}
\end{equation*}
This defines the category of $T$-algebras.  Notice that, for any
$T$-algebra $A$, there is an induced $T$-algebra structure on $T A$
given by $T\sigma_A\colon T (T A) \to T A$, turning $\sigma_A$ into a
morphism of algebras.  An initial $T$-algebra is called a
\emph{(least) fixpoint} of $T$.  Given one such fixpoint $U$ and any
other $T$-algebra $A$ we denote by $\fold_A\colon U \to A$ the induced
initial morphism.  We observe that $\sigma_U$ is an isomorphism whose
inverse is $\fold_{TU}$ since
\begin{math}
   \sigma_U \cdot \fold_{TU} = 1_U
\end{math}
by the universal property of $U$ and from the naturality of $\fold$
follows that the diagram
\begin{equation*}
  \xymatrix{
    T U \ar[d]_{\sigma_U}\ar[rrd]^{1_{TU}}\ar[rr]^{T\,\fold_{TU}} &&
      T (T U)\ar[d]^{T\sigma_U}\\
    U \ar[rr]_{\fold_{TU}} && TU}
\end{equation*}
is commutative.

Let us show how this general framework can work in the case of
(polymorphic) lists.

\begin{example}
  Take $\CC=\Set\to\Set$ the category of endofunctors of $\Set$
  and consider the functor $T\colon (\Set\to\Set) \to (\Set\to\Set)$
  defined by
  \begin{equation*}
    T(F) \coloneq X \mapsto * + X \times F X.
  \end{equation*}
  The least fix point of $T$ is (the underlying functor of) the monad
  of lists $L$ introduced in section \ref{sec:monads}.  The
  $T$-algebra structure $*+X\times L X = T L \simeq L$ gives the
  constructors ($\mathsf{nil}$, $\mathsf{cons}$) and the corresponding
  destructors.  We would like to recognise this structural isomorphism
  as an $L$-linear morphism.  Unfortunately, we do not have on $TL$ a
  structure of left $L$-module corresponding to our expectation
  (notice that the identity functor is not an $L$-module in a natural
  way).  We will explain in section \ref{sec:monads-over-types} how
  this phenomenon can be considered a consequence of the lack of
  typing.
\end{example}

\section{Monads over sets}
\label{sec:monads-over-sets}

In this section we consider more examples of linear morphisms over
monads on the category of sets.

\subsection{Untyped Syntactic Lambda Calculus}
\label{sec:untyp-lambda-calc}

Consider the functor $T\coloneq (\Set \to \Set) \to (\Set \to \Set)$
given by
\begin{equation*}
  T F \colon X \mapsto X + F X\times F X + F' X
\end{equation*}
where $F'$ denotes the derived functor $X\mapsto F(X+*)$.  It can be
shown that $T$ possesses a least fixpoint that we denote by $\LC$
($\LC$ standing for $\lambda$-calculus, cfr.\ the example in section
\ref{sec:monads}).  We consider $\LC(X)$ as the set of $\lambda$-terms with
free variables taken from $X$ (see also \cite{BPdebruijn}).  In fact,
the structural morphism $T\LC \to \LC$ gives the familiar constructors
for $\lambda$-calculus in the \emph{locally nameless} encoding,
namely, the natural transformations
\begin{equation*}
  \var \colon I \to \LC, \quad
  \app \colon \LC \times \LC \to \LC, \quad
  \abs \colon \LC' \to \LC.
\end{equation*}
As already observed, the substitution (instantiation) of free
variables gives a monad structure on $\LC$ where $\var$ is the unit.

We would like to express that these constructors are well behaved with
respect to substitution.  Again, as in the case of lists, $T \LC$ has
no natural structure of left $\LC$-module.  However, we can consider
the functor $T$ as built of two parts $T F = I + T_0 F$ where $T_0 F
\coloneq F\times F + F'$ (in other words we are tackling apart $\var$,
the unit of the monad, from the other two constructors $\app$ and
$\abs$).  Now $T_0 \LC$ is a left $\LC$-module and we can observe that
the algorithm of substitution is carried precisely in such a way that
the induced morphism
\begin{equation*}
  \app,\abs \colon T_0 \LC \to \LC
\end{equation*}
is $\LC$-linear or, equivalently, the natural transformations
$\app\colon \LC\times\LC\to \LC$ and $\abs\colon \LC' \to \LC$ are
$\LC$-linear.  To make the idea clearer, we reproduce a short piece of
code in the $Haskell$ programming language which implements the
algorithm of substitution.
\begin{small}
\begin{verbatim}
  module LC where
  import Monad (liftM)
\end{verbatim}
\begin{verbatim}
  data LC a = Var a | App (LC a) (LC a) | Abs (LC (Maybe a))
\end{verbatim}
\begin{verbatim}
  instance Monad LC where
      return = Var
      Var x >>= f = f x
      App x y >>= f = App (x >>= f) (y >>= f)
      Abs x >>= f = Abs (x `mbind` f)

  mbind :: LC (Maybe a) -> (a -> LC b) -> LC (Maybe b)
  mbind x f = x >>= maybe (Var Nothing) (liftM Just . f)
\end{verbatim}
\end{small}
In the above code, $\tt{mbind}$ constitutes the left $\LC$-module
structure on $\LC'$.  It is now evident that the recursive calls in the
definition of \texttt{(>>=)} are exactly those given by the linearity
of $\app$ and $\abs$.

We can go further and try to make the linearity more explicit in the
syntactic definition of $\lambda$-calculus.  This can be done as
follows.
\begin{theorem}
  Consider the category $\Mon^{T_0}$ where objects are monads $R$ over
  sets endowed with a $R$-linear morphism $T_0R\to R$ while arrows are
  given by commutative diagrams
  \begin{equation*}
    \xymatrix{
      T_0 R \ar[d] \ar[r]^{T_0 f} &  f^*T_0 P\ar[d] \\
      R \ar[r]^f & f^*P}
  \end{equation*}
  where all morphisms are $R$-linear (we are using implicitly the fact
  that the base change functor commutes with derivation and products).
  The monad $\LC$ is initial in $\Mon^{T_0}$.
\end{theorem}

In fact, the previous theorem can be generalized as follows
(interested readers may also want to look at other works on higher
order abstract syntax, e.g., \cite{FPT,H,PG} see also our
\cite{Algebraicity}).  Let $R$ be a monad over $\Set$.  We define an
\emph{arity} to be a list of nonnegative integers.  We denote by
$\NN^*$ the set of arities.  For each arity $(a_1, \dots, a_r)$, and
for any $R$-module $M$, we define the $R$-module $T^a M$ by
\begin{equation*}
  T^a M = M^{(a_1)}\times \cdots \times M^{(a_r)},
\end{equation*}
where $M^{(n)}$ denotes the $n$-th derivative of $M$,
and we say that a linear morphism $T^a R \to R$ is a
$R$-\emph{representation} of $a$ (or a representation of $a$ in $R$).
For instance, the $\app$ and
$\abs$ constructors are $\LC$-representations of the arities $(0,0)$
and $(1)$ respectively.

Next, we consider \emph{signatures} which are family of arities.  For
each signature $\Sigma \colon I \to \NN^*$, and for any $R$-module $M$,
 we define the $R$-module $T^\Sigma M$  by
\begin{equation*}
  T^\Sigma M= \sum_{i\in I} T^{\Sigma_i} M
\end{equation*}
and we say that a linear morphism $T^\Sigma R \to R$ is a
$R$-\emph{representation} of $\Sigma$ (or a representation of $\Sigma$ in $R$).
Altogether $\app$ and $\abs$ give a $\LC$-representation of the
signature $((0,0),(1))$.

As in the special case of the $\lambda$-calculus, representations of a
given signature $\Sigma$ form a category.

\begin{theorem}
  \label{thm:signature-representation-initial}
  For any signature $\Sigma$, the category of $\Sigma$-representations
  has an initial object.
\end{theorem}

\subsection{Untyped Semantic Lambda Calculus}
\label{sec:untyp-sem-lambda-calc}

For any set $X$, consider the equivalence relation
$\equiv_{\beta\eta}$ on $\LC(X)$ given by the reflexive symmetric
transitive closure of $\beta$ and $\eta$ conversions and define
$\Lambda(X) \coloneq \LC(X)/\equiv_{\beta\eta}$.  It can be shown that
$\equiv_{\beta\eta}$ is compatible with the structure of $\LC$ so
$\Lambda$ has a structure of monad, the projection $\LC\to\Lambda$ is
a morphism of monads, and we have an induced morphism $T_0\Lambda \to
\Lambda$ which is $\Lambda$-linear.

Now the key fact is that the abstraction $\abs\colon \Lambda' \to
\Lambda$ is a linear isomorphism!  In fact, it is easy to
construct its inverse $\app_1 \colon \Lambda \to \Lambda'$:
\begin{equation*}
  \app_1\, x = \app(\hat x,*)
\end{equation*}
where $x\mapsto \hat x$ denotes the natural inclusion $\Lambda \to
\Lambda'$.
The equation
\begin{equation*}
  \abs\cdot\app_1 = 1_\Lambda
\end{equation*}
clearly corresponds to the $\eta$-rule while the other equation
\begin{equation*}
  \app_1\cdot\abs = 1_{\Lambda'}
\end{equation*}
can be considered the ultimate formulation of the $\beta$-rule.  In
fact, there is a more classical formulation of the $\beta$-rule which
can be stated as the commutativity of the diagram
\begin{equation*}
 \xymatrix{
   \Lambda'\times \Lambda \ar[r]^{\abs\times\Lambda}\ar[dr]_{\mathsf{subst}} &
   \Lambda\times \Lambda\ar[d]^\app \\
   {} & \Lambda}
\end{equation*}

Again, we can present this situation from a more syntactical point of
view.  For this, consider the category of \emph{exponential monads}:
an exponential monad is a monad $R$ endowed with a $R$-linear
isomorphism with its derivative $\exp_R\colon R' \to R$.  An arrow is
a monad morphism $f$ such that
\begin{equation*}
  \xymatrix{
    R' \ar[d]_{\exp_R} \ar[r]^{f'} & f^*P'\ar[d]^{\exp_P} \\
    R \ar[r]^f & f^*P}
\end{equation*}
is a commutative diagram of $R$-modules (we are implicitly using the
commutativity of base change with derivation).
\begin{theorem}
  \label{thm:lc-initial}
  The monad $\Lambda$ is initial in the category of exponential
  monads.
\end{theorem}
We have developed a formal proof of the above theorem in the Coq proof
assistant \cite{Coq} which is discussed in the appendix.

\section{Monads over types}
\label{sec:monads-over-types}

So far we mostly considered examples of monads and modules on the
category $\CC=\Set$ of small sets.  Other interesting phenomena can be
captured by taking into account monads and modules on other
categories.  In this section we consider the case $\CC=\Set_\tau$ the
category of sets fibered over a fixed set $\tau$.  This is the
category given by maps $\phi_X\colon X\to \tau$, called $\tau$-sets,
where arrows $\langle X,\phi_X\rangle \to\langle Y,\phi_Y\rangle$ are
given by maps $f\colon X \to Y$ which commute with the structural
morphisms, i.e., $\phi_Y\cdot f = \phi_X$.  For each $t\in\tau$ and
each $\tau$-set $X$, we denote by $X_\tau \coloneq \phi_X^{-1}(t)$ the
preimage of $t$ in $t$.  We regard $\tau$ as a ``set of types'' and
the fibers $X_t$ as a set of ``terms of type $t$''.

\subsection{Typed lists}
\label{sec:typed-lists}

Here we show how, in the context of typed lists, the constructors
$\nil$ and $\cons$ may appear as linear.  To this effect, we introduce
a distinction between the base type $*$, the type of lists
$\mathsf{list}\, *$, the type of lists of lists, etc.  Thus we take
$\tau=\NN$ the inductive set of types generated by the grammar $\tau =
*\, |\,\mathsf{list}\,\tau$, and consider the category $\Set_\tau$.

For each $t \in \tau$ we define $\EuScript{L}_t: \Set_\tau \to \Set$
by setting $\EuScript{L}_t(X)$ to be the set of terms of type $t$
built out from (typed) variables in $X$ by adding, as usual, terms
obtained through the $\mathsf{nil}$ and $\mathsf{cons}$ constructions.
By glueing these $\EuScript{L}_t$ together, we obtain an endofunctor
$\EuScript{L}$ in $\Set_\tau$.  It is easily seen to be a monad (the
present structure of monad has nothing to do with flattening).

For each $t \in \tau$, $\EuScript{L}_t$ is a left
$\EuScript{L}$-module (by Example 7).  The $\nil$ and $\cons$
constructors constitute a family of natural transformations
parametrized by $t\in\tau$
\begin{equation*}
 \nil_t \colon * \longrightarrow \EuScript{L}_{\mathsf{list}\, t}, \qquad
 \cons_t \colon \EuScript{L}_t \times \EuScript{L}_{\mathsf{list}\, t} \longrightarrow
 \EuScript{L}_{\mathsf{list}\, t}.
\end{equation*}
Hence we have here examples of heterogeneous modules since $*$,
$\EuScript{L}_t$ and $\EuScript{L}_{\mathsf{list}\, t}$ are
$\EuScript{L}$-modules in $\Set$. And $\nil_t$ and $\cons_t$ are
easily seen to be morphisms among these modules.

We may also want to glue for instance these $\cons_t$ into a single
$\cons$.  For this, we need shifts.  Given $X\in\Set_\tau$ we have the
associated \emph{shifts} $X[n]$ which are obtained by adding $n$ to
the structural map $X\to \NN$.  The shift $(\cdot)[n] \colon X\mapsto
X[n]$ gives an endofunctor over $\Set_\tau$.  Given a functor $F$ from
any category to the category $\Set_\tau$, we consider the
\emph{shifted functors} $F[n]$ obtained as composition of $F$ followed
by $(\cdot)[n]$.  From the remarks of section \ref{sec:modules}, it
follows at once that if $F$ is an $\EuScript{L}$-module, then so is
$F[n]$.  With these notations, glueing yields
\begin{equation*}
 \nil \colon *[1] \longrightarrow \EuScript{L},\qquad
 \cons \colon  \EuScript{L}[1]\times \EuScript{L} \longrightarrow \EuScript{L}
\end{equation*}
where $*$ denotes the final functor in the category of endofunctors of
$\Set_\tau$.  Again $\nil$ and $\cons$ are easily checked to be
$\EuScript{L}$-linear.

\subsection{Simply Typed Lambda Calculus}
\label{sec:styped-lambda-calc}

Our second example of typed monad is the simply-typed
$\lambda$-calculus.  We denote by $\tau$ the set of simple types $\tau
\coloneq *\ |\ \tau \to \tau$.  Following \cite{JZ}, we consider the
syntactic typed $\lambda$-calculus as an assignment $V \mapsto
\LC_\tau(V)$, where $V= \sum_{t \in \tau} (V_t) $ is a (variable) set
(of typed variables) while
\begin{equation*}
  \LC_\tau(V)= \sum_{t \in \tau}(\LC_\tau(V))_t
\end{equation*}
is the set of typed $\lambda$-terms (modulo $\alpha$-conversion) built
on free variables taken in $V$.

Given a type $t$ we set $\LC_t(X) \coloneq (\LC_\tau(X))_t$ which
gives a functor over $\tau$-sets, which is equipped with substitution,
turning it into a (heterogeneous) left module over $\LC_\tau$.  And
given two types $s, t$, we have
\begin{equation*}
  \app_{s,t} \colon \LC_{s \to t} \times \LC_s \longrightarrow \LC_t
\end{equation*}
which is linear.

For the $\abs$ construction, we need a notion of partial derivative
for a module.  For a left module $M$ over $\tau$-sets, and a type $t
\in \tau$, we set
\begin{equation*}
  \delta_t M (V) \coloneq M(V+*_t)
\end{equation*}
where $V+*_t$ is obtained from $V$ by adding one element with type
$t$.  It is easily checked how $\delta_t M$ is again a left module.
Now, given two types $s$ and $t$, it turns out that
\begin{equation*}
  \abs_{s,t} \colon \delta_s \LC_t \longrightarrow \LC_{s \to t}
\end{equation*}
is linear.

As in the untyped case, we can consider the functor $\Lambda_\tau$
obtained by quotienting modulo $\beta\eta$ conversion.  This is again
a monad over the category of $\tau$-sets and the natural quotient
transformation $\LC_\tau \to \Lambda_\tau$ is a morphism of monads.
For this semantic monad, the above left module morphisms induce
semantic counterparts:
\begin{math}
 \app_{s,t} \colon \Lambda_{s \to t} \times \Lambda_s \longrightarrow \Lambda_t
\end{math}
and
\begin{math}
 \abs_{s,t} \colon \delta_s \Lambda_t \longrightarrow \Lambda_{s \to t}
\end{math}.

Here we need a new notion of arity and signature, which we will
introduce in some future work, in order to state and prove a typed
counterpart of our theorem \ref{thm:signature-representation-initial}.
For a typed counterpart of our theorem \ref{thm:lc-initial}, see
\cite{JZ}.

\subsection{Typed Lambda Calculus}
\label{sec:typed-lambda-calc}

Our final example of typed monad is just a glance to more general
typed $\lambda$-calculi.  The point here is that the set of types is
no more fixed.  Thus our monads take place in the category $\Cat{Fam}$
of set families: an object in $\Cat{Fam}$ is an application $p \colon
I \to \Set$ while a morphism $m \colon p \to p'$ is a pair $(m_0,
m_1)$ with $m_0 \colon I \to I'$, and $m_1\colon (i:I) p(i) \to
p'(m_0(i))$.  We say that $I$ is the set of types of $p \colon I \to
\Set$.  From $\Cat{Fam}$ there are two forgetful functors $T,
\mathop{Tot} \colon \Cat{Fam} \to \Set$ where $T(p\colon I \to \Set)
\coloneq I$ and $\mathop{Tot}(p) \coloneq \amalg_{i \in T(p)}p(i)$,
and a natural transformation $\mathop{proj} \colon \mathop{Tot} \to
T$, defined by $\mathop{proj} (p)=p$ in the obvious sense.  Given a
monad $R$ on $\Cat{Fam}$, we thus have a morphism of $R$-modules
$\mathop{proj}_R: \mathop{Tot} \circ R \to T \circ R.$

We need also two {\em may-be} monads on $\Cat{Fam}$: the first one $F
\mapsto F^*$ adds one (empty) type ($tnew$) to $F$, while the second
one, $F \mapsto F^{*/*}$ adds one type ($tnew$) with one element
($new$).  Given a monad $R$ on $\Cat{Fam}$, we thus have two
``derived'' $R$-modules: $R^*:= F \mapsto R(F^*)$ and $R^{*/*}
\coloneq F \mapsto R(F^{*/*})$

Now when should we say that $R$ is a lambda-calculus in this context?
At least we should have a module morphism $arrow: (T \circ R)^2 \to T \circ R.$
and a module morphism for abstraction, $abs: R^{*/*} \to R^*$ (the ``arity'' for application is not so simple). We hope this example shows the need for new notions of arity and signature, as well as the new room opened by modules for such concepts.

\section{Monads over preordered sets}
\label{sec:monads-over-preord}

Our last example is about monads and modules over the category of
preordered sets (sets with a reflexive and transitive binary
relation).  Preordering is used here to model the relation
$\stackrel{\beta\eta}\longrightarrow_*$ generated by the reflexive and
transitive closure of the $\beta$ and $\eta$ conversions.  In fact,
the construction given in this section can be considered a refinement
of those of section \ref{sec:untyp-lambda-calc} where we used the
reflexive, symmetric and transitive closure $\equiv_{\beta\eta}$.

Let us consider again the monad $\LC$ of $\lambda$-terms.  Given a
preordered set $X$, we consider the preordering on $\LC(X)$ given by
the rules
\begin{eqnarray*}
  x\leq y &\implies& \var\,x\leq\var\,y, \\
  S\leq S' \wedge T\leq T' &\implies& \app(S,T)\leq \app(S',T'), \\
  T\leq T' &\implies& \abs(T)\leq \abs(T'), \\
  T\longrightarrow_{\beta\eta} T' &\implies& T \leq T'.
\end{eqnarray*}

It is not hard to verify that with this new structure $\LC$ is now a
monad over preordered sets.  It turns out that the $\app$ and $\abs$
constructions are still $\LC$-linear with respect to this richer
structure.

\section{Conclusions and related works}
\label{sec:conclusions}

We have introduced the notion of module over a monad, and more
importantly the notion of linearity for transformations among such
modules and we have tried to show that this notion is ubiquitous as
soon as syntax and semantics are concerned.  Our thesis is that the
point of view of modules opens some new room for initial algebra
semantics, as we sketched for typed $\lambda$-calculus (see also
\cite{Algebraicity}).

The idea that the notion of monad is suited for modelling substitution
concerning syntax (and semantics) has been retained by many recent
contributions concerned with syntax (see e.g.\
\cite{BPfold,GU03,MU03}) although some other settings have been
considered.  Notably in \cite{FPT} the authors work within a setting
roughly based on operads (although they do not write this word down;
the definition of operad is on Wikipedia; operads and monads are not
too far from each other).  As they mention, their approach is, to some
extent, equivalent to an approach through monads.  It has been both
applied e.g.\ in \cite{Power-Tanaka-pseudo} and generalized e.g.\ in
\cite{Power-Tanaka-unified}.  Another approach to syntax with
bindings, initiated by Gabbay and Pitts \cite{PG}, relies on a
systematic consideration of freshness, an issue which is definitely
ignored in the monadic or operadic approach.

While the notion of module over a monad has been essentially ignored
till now, the notion of module over an operad has been introduced more
than ten years ago, and has been incidentally considered in the
context of semantics, as we already mentioned in our introduction.

\section{Appendix: Formal proof of theorem \ref{thm:lc-initial}}
\label{sec:appendix}

In this section we present our formal proof of theorem
\ref{thm:lc-initial} in the Coq proof assistant \cite{Coq}.  We recall
the statement of the theorem
\begin{quote}
  The monad $\Lambda$ of semantic untyped $\lambda$-calculus is an
  initial object in the category of exponential monads.
\end{quote}
We include here only a small fraction of the code without proofs.  The
full sources can be found at \verb+http://www.math.unifi.it/~maggesi+.

\subsection{Structure of the formalisation}
\label{sec:struct-form}

The structure of our proof can be outlined in the following four major parts:
(1) axioms and support library;
(2) formalisation of monads, modules and exponential monads;
(3) formalisation of syntactic and semantic $\lambda$-calculus;
(4) the main theorem.

The second and third part are independent of each other.  As for what
this paper is concerned, the first part (files Misc.v, Congr.v) can be
considered as an extension of the Coq system for practical purposes.
This part contains some meta-logical material (tactics and notations)
and declares the following axioms: functional choice, proof
irrelevance, dependent extensionality.  We include here their
declarations:
\begin{small}
\begin{verbatim}
  Axiom functional_choice : forall (A B : Type) (R : A -> B -> Prop),
    (forall x : A,  exists y : B, R x y) -> exists f : A -> B, (forall x : A, R x (f x)).
  Axiom proof_irrelevance : forall (A : Prop) (H1 H2 : A), H1 = H2.
  Axiom extens_dep : forall (X : Type) (T : X -> Type) (f g : forall x : X, T x),
    (forall x : X, f x = g x) -> f = g.
\end{verbatim}
\end{small}
Moreover, we use an axiomatic definition of quotient types (file
Quot.v) to construct semantic $\lambda$-calculus as quotient of
syntactic $\lambda$-calculus.

\subsection{Formalisation of monads and modules}
\label{sec:form-monads-modul}

After the preliminary material, our formalisation opens the theory of
monads and (left) modules (files Monad.v, Mod.v, Derived\_Mod.v).  This
is constructed starting from a rather straightforward translation of
the Haskell monad library.  As an example, we report here our
definitions of monads and modules in the Coq syntax.
\begin{small}
\begin{verbatim}
  Record Monad : Type := {
    monad_carrier :> Set -> Set;
    bind : forall X Y : Set, (X -> monad_carrier Y) -> monad_carrier X -> monad_carrier Y;
    unit : forall X : Set, X -> monad_carrier X;
    bind_bind : forall (X Y Z : Set) (f : X -> monad_carrier Y) (g : Y -> monad_carrier Z)
      (x : monad_carrier X),
      bind Y Z g (bind X Y f x) = bind X Z (fun u => bind Y Z g (f u)) x;
    bind_unit : forall (X Y : Set) (f : X -> monad_carrier Y) (x : X),
      bind X Y f (unit X x) = f x;
    unit_bind : forall (X : Set) (x : monad_carrier X), bind X X (unit X) x = x
  }.
  Notation "x >>= f" := (@bind _ _ _ f x).
\end{verbatim}
\end{small}
\begin{small}
\begin{verbatim}
  Record Mod (U : Monad) : Type := {
    mod_carrier :> Set -> Set;
    mbind : forall (X Y: Set) (f : X -> U Y) (x : mod_carrier X), mod_carrier Y;
    mbind_mbind : forall (X Y Z : Set) (f : X -> U Y) (g : Y -> U Z) (x : mod_carrier X),
      mbind Y Z g (mbind X Y f x) = mbind X Z (fun u => f u >>= g) x;
    unit_mbind : forall (X : Set) (x : mod_carrier X), mbind X X (@unit U X) x = x
  }.
  Notation "x >>>= f" := (@mbind _ _ _ _ f x).
\end{verbatim}
\end{small}
The library also includes the definition of morphism of monads and
modules and other related categorical material.  Other definitions
which are specific to our objective are those of derived module and
exponential monad.  The latter reads as follows:
\begin{small}
\begin{verbatim}
  Record ExpMonad : Type := {
    exp_monad :> Monad;
    exp_abs : Mod_Hom (Derived_Mod exp_monad) exp_monad;
    exp_app : Mod_Hom exp_monad (Derived_Mod exp_monad);
    exp_eta : forall X (x : exp_monad X), exp_abs _ (exp_app _ x) = x;
    exp_beta : forall X (x : Derived_Mod exp_monad X), exp_app _ (exp_abs _ x) = x
  }.
\end{verbatim}
\end{small}
and comes with its associated notion of morphism:
\begin{small}
\begin{verbatim}
  Record ExpMonad_Hom (M N : ExpMonad) : Type := {
    expmonad_hom :> Monad_Hom M N;
    expmonad_hom_app : forall X (x : M X),
      expmonad_hom _ (exp_app M _ x) = exp_app N _ (expmonad_hom _ x);
    expmonad_hom_abs : forall X (x : Derived_Mod M X),
      expmonad_hom _ (exp_abs M _ x) = exp_abs N _ (expmonad_hom _ x)
  }.
\end{verbatim}
\end{small}

\subsection{Formalisation of the $\lambda$-calculus}
\label{sec:form-lambda-calc}

This part contains the definition of syntactic and semantic
$\lambda$-calculus (files Slc.v and Lc.v respectively).  We use nested
datatypes to encode $\lambda$-terms in the Calculus of (Co)Inductive
Constructions as already shown in the Haskell fragment of section
\ref{sec:untyp-lambda-calc} for which we report below the equivalent
Coq code.  Notice that this encoding can be considered a typeful
variant of the well-known de Bruijn encoding \cite{BPdebruijn}.  As
the de Bruijn encoding, it represents $\lambda$-terms modulo
$\alpha$-conversion.
\begin{small}
\begin{verbatim}
  Inductive term (X : Set) : Set := var : X -> term X
                                  | app : term X -> term X -> term X
                                  | abs : term (option X) -> term X.
\end{verbatim}
\end{small}
\begin{small}
\begin{verbatim}
  Fixpoint fct (X Y : Set) (f : X -> Y) (x : term X) { struct x } : term Y :=
    match x with var a => var (f a)
               | app x y => app (x //- f) (y //- f)
               | abs x => abs (x //- (optmap f)) end
  where "x //- f" := (@fct _ _ f x).
\end{verbatim}
\end{small}
\begin{small}
\begin{verbatim}
  Definition shift X (x : term X) : term (option X) := x //- @Some X.
\end{verbatim}
\end{small}
\begin{small}
\begin{verbatim}
  Definition comm (X Y : Set) (f : X -> term Y) (x : option X) : term (option Y) :=
    match x with Some a => shift (f a) | None => var None end.
\end{verbatim}
\end{small}
\begin{small}
\begin{verbatim}
  Fixpoint subst (X Y : Set) (f : X -> term Y) (x : term X) { struct x } : term Y :=
    match x with var x => f x
               | app x y => app (x //= f) (y //= f)
               | abs x => abs (x //= comm f) end
  where "x //= f" := (@subst _ _ f x).
\end{verbatim}
\end{small}
Once the basic definitions are settled, we prove a series of basic
lemmas which includes the associativity of substitution
\begin{small}
\begin{verbatim}
  Lemma subst_subst : forall (X Y Z : Set) (f : X -> term Y) (g : Y -> term Z) (x : term X),
    x //= f //= g = x //= fun u => f u //= g.
\end{verbatim}
\end{small}
which is the most important ingredient to prove that the $\lambda$-calculus
is a monad.  Finally, we introduce the beta-eta equivalence relation
on lambda terms
\begin{small}
\begin{verbatim}
  Inductive lcr (X : Set) : term X -> term X -> Prop :=
    | lcr_var : forall a : X, var a == var a
    | lcr_app : forall x1 x2 y1 y2 : term X, x1 == x2 -> y1 == y2 -> app x1 y1 == app x2 y2
    | lcr_abs : forall x y : term (option X), x == y -> abs x == abs y
    | lcr_beta : forall x y : term X, Beta x y -> x == y
    | lcr_eta : forall x : term X, abs (app1 x) == x
    | lcr_sym : forall x y : term X, y == x -> x == y
    | lcr_trs : forall x y z : term X, lcr x y -> lcr y z -> lcr x z
  where "x == y" := (@lcr _ x y).
\end{verbatim}
\end{small}
and prove some compatibility lemmas for constructors and other
operations.  The compatibility of substitution is stated as follows:
\begin{small}
\begin{verbatim}
  Lemma lcr_subst : forall (X Y : Set) (f g : X -> term Y) (x y : term X),
    (forall u, f u == g u) -> x == y -> x //= f == y //= g.
\end{verbatim}
\end{small}

\subsection{Proof of the main theorem}
\label{sec:proof-main-theorem}

The fourth and last part summarises the results proved in the other
parts and proves the main theorem.  It starts by glueing together the
previous two sections by proving that our definitions of syntactic and
semantic lambda calculus provides indeed two monads, denoted
$\mathtt{SLC}$ and $\mathtt{LC}$ respectively, and by showing that the
two morphisms $\abs$ and $\app_1$ constitute morphisms of modules:
\begin{small}
\begin{verbatim}
  Definition SLC : Monad := Build_Monad term subst var subst_subst subst_var var_subst.
\end{verbatim}
\end{small}
\begin{small}
\begin{verbatim}
  Definition LC : Monad :=
    Build_Monad lc lc_subst lc_var lc_subst_assoc lc_subst_var lc_var_subst.
\end{verbatim}
\end{small}
\begin{small}
\begin{verbatim}
  Let abs_hom : Mod_Hom (Derived_Mod LC) LC :=
    Build_Mod_Hom (Derived_Mod LC) LC lc_abs lc_abs_hom.
\end{verbatim}
\end{small}
\begin{small}
\begin{verbatim}
  Let app1_hom : Mod_Hom LC (Derived_Mod LC) :=
    Build_Mod_Hom LC (Derived_Mod LC) lc_app1 lc_app1_hom.
\end{verbatim}
\end{small}
One more glueing step is the proof that $\mathtt{LC}$ is an exponential
monad, which is stated in Coq through the following definition:
\begin{small}
\begin{verbatim}
  Definition ELC : ExpMonad := Build_ExpMonad abs_hom app1_hom lc_eta lc_beta.
\end{verbatim}
\end{small}
Next comes the construction of the initial morphism which is initially
defined as a fixpoint on terms.
\begin{small}
\begin{verbatim}
  Variable M : ExpMonad.
  Fixpoint iota_fix X (x : term X) { struct x } : M X :=
    match x with var a => unit M a
               | app x y => exp_app M _ (iota_fix x) >>= default (@unit M X) (iota_fix y)
               | abs x => exp_abs M _ (iota_fix x) end.
\end{verbatim}
\end{small}
Then we prove that $\mathtt{iota\_fix}$ is compatible with the $\beta\eta$ equivalence relation and thus it factors through the monad $\mathtt{LC}$.
\begin{small}
\begin{verbatim}
  Let iota X : lc X -> M X := lc_factor (@iota_fix X) (@iota_fix_wd X).
\end{verbatim}
\end{small}
The construction of the initial morphism ends with the verification
that it is actually a morphism of exponential monads.
\begin{small}
\begin{verbatim}
  Let iota_monad : Monad_Hom LC M := Build_Monad_Hom LC M iota iota_subst iota_var.
\end{verbatim}
\end{small}
\begin{small}
\begin{verbatim}
  Let exp_iota : ExpMonad_Hom ELC M :=
    Build_ExpMonad_Hom ELC M iota_monad iota_app1 iota_abs.
\end{verbatim}
\end{small}
Finally, we prove that $\mathtt{iota\_monad}$ is unique.
\begin{small}
\begin{verbatim}
  Theorem iota_unique : forall j : ExpMonad_Hom ELC M, j = exp_iota.
\end{verbatim}
\end{small}
The Coq terms $\mathtt{ELC}$, $\mathtt{iota\_monad}$ and
$\mathtt{iota\_unique}$ altogether form the formal proof of the
initiality of the monad $\mathtt{LC}$ in the category of exponential
monads.

\bibliographystyle{plain}
\bibliography{syntax}

\begin{thebibliography}{10}

\bibitem{Alt-Reus}
Thorsten Altenkirch and Bernhard Reus.
\newblock Monadic presentations of lambda terms using generalized inductive
  types.
\newblock In {\em {CSL}}, pages 453--468, 1999.

\bibitem{BPfold}
Richard Bird and Ross Paterson.
\newblock Generalised folds for nested datatypes.
\newblock {\em Formal Aspects of Computing}, 11(2):200--222, 1999.

\bibitem{BPdebruijn}
Richard~S. Bird and Ross Paterson.
\newblock De {B}ruijn notation as a nested datatype.
\newblock {\em Journal of Functional Programming}, 9(1):77--91, 1999.

\bibitem{MR2140994}
Michael Ching.
\newblock Bar constructions for topological operads and the {G}oodwillie
  derivatives of the identity.
\newblock {\em Geom. Topol.}, 9:833--933 (electronic), 2005.

\bibitem{Coq}
The {Coq Proof Assistant}.
\newblock http://coq.inria.fr.

\bibitem{FPT}
Marcelo Fiore, Gordon Plotkin, and Daniele Turi.
\newblock Abstract syntax and variable binding (extended abstract).
\newblock In {\em 14th Symposium on Logic in Computer Science (Trento, 1999)},
  pages 193--202. IEEE Computer Soc., Los Alamitos, CA, 1999.

\bibitem{Fiore06}
Marcelo~P. Fiore.
\newblock On the structure of substitution.
\newblock Invited address for the 22nd Mathematical Foundations of Programming
  Semantics Conf. (MFPS XXII), 2006.
\newblock DISI, University of Genova (Italy).

\bibitem{FT}
Marcelo~P. Fiore and Daniele Turi.
\newblock Semantics of name and value passing.
\newblock In {\em Logic in Computer Science}, pages 93--104, 2001.

\bibitem{MR2066499}
Benoit Fresse.
\newblock Koszul duality of operads and homology of partition posets.
\newblock In {\em Homotopy theory: relations with algebraic geometry, group
  cohomology, and algebraic $K$-theory}, volume 346 of {\em Contemp. Math.},
  pages 115--215. Amer. Math. Soc., Providence, RI, 2004.

\bibitem{PG}
Murdoch Gabbay and Andrew Pitts.
\newblock A new approach to abstract syntax involving binders.
\newblock In {\em 14th Symposium on Logic in Computer Science (Trento, 1999)},
  pages 214--224. IEEE Computer Soc., Los Alamitos, CA, 1999.

\bibitem{GU03}
Neil Ghani and Tarmo Uustalu.
\newblock Explicit substitutions and higher-order syntax.
\newblock In {\em MERLIN '03: Proceedings of the 2003 ACM SIGPLAN workshop on
  Mechanized reasoning about languages with variable binding}, pages 1--7, New
  York, NY, USA, 2003. ACM Press.

\bibitem{Algebraicity}
Andr\'e Hirschowitz and Marco Maggesi.
\newblock The algebraicity of the lambda-calculus.
\newblock arXiv:math/0607427v1, 2007.

\bibitem{H}
Martin Hofmann.
\newblock Semantical analysis of higher-order abstract syntax.
\newblock In {\em 14th Symposium on Logic in Computer Science (Trento, 1999)},
  pages 204--213. IEEE Computer Soc., Los Alamitos, CA, 1999.

\bibitem{arXiv:math/0607427v1}
Muriel Livernet.
\newblock From left modules to algebras over an operad: application to
  combinatorial {H}opf algebras.
\newblock arXiv:math/0607427v1, 2006.

\bibitem{maclane}
Saunders Mac~Lane.
\newblock {\em Categories for the working mathematician}, volume~5 of {\em
  Graduate Texts in Mathematics}.
\newblock Springer-Verlag, New York, second edition, 1998.

\bibitem{arXiv:hep-th/9411208v1}
Martin Markl.
\newblock Models for operads.
\newblock arXiv:hep-th/9411208v1, 1994.

\bibitem{arXiv:hep-th/9608067v1}
Martin Markl.
\newblock A compactification of the real configuration space as an operadic
  completion.
\newblock arXiv:hep-th/9608067v1, 1996.

\bibitem{MU03}
Ralph Matthes and Tarmo Uustalu.
\newblock Substitution in non-wellfounded syntax with variable binding.
\newblock {\em Theor. Comput. Sci.}, 327(1-2):155--174, 2004.

\bibitem{smirnov86}
V.~A. Smirnov.
\newblock Homotopy theory of coalgebras.
\newblock {\em Math. USSR Izv,}, 27:575--592, 1986.

\bibitem{Power-Tanaka-unified}
Miki Tanaka and John Power.
\newblock A unified category-theoretic formulation of typed binding signatures.
\newblock In {\em MERLIN '05: Proceedings of the 3rd ACM SIGPLAN workshop on
  Mechanized reasoning about languages with variable binding}, pages 13--24,
  New York, NY, USA, 2005. ACM Press.

\bibitem{Power-Tanaka-pseudo}
Miki Tanaka and John Power.
\newblock Pseudo-distributive laws and axiomatics for variable binding.
\newblock {\em Higher Order Symbol. Comput.}, 19(2-3):305--337, 2006.

\bibitem{JZ}
Julianna Zsid\'o.
\newblock Le lambda calcul vu comme monade initiale.
\newblock Master's thesis, Universit\'e de Nice -- Laboratoire
  J.~A.~Dieudonn\'e, 2005/06.
\newblock M\'emoire de Recherche -- master 2.

\end{thebibliography}

\end{document}